\documentclass[aps, amsfonts, amssymb, amsmath, reprint, nofootinbib, twoside, twocolumn, superscriptaddress, url]{revtex4-2}

\usepackage[english]{babel}
\usepackage[utf8]{inputenc}
\usepackage{times} 

\usepackage[colorinlistoftodos, color=green!40, prependcaption]{todonotes}
\usepackage{graphicx}
\usepackage{soul}
\usepackage{xfrac}
\usepackage{bm}
\usepackage{float}
\usepackage{newfloat,algcompatible}
\usepackage{ulem}
\graphicspath{{Figures/}}
\usepackage{color}
\usepackage{makecell}

\usepackage{url}
\usepackage{caption}
\usepackage{subcaption}
\usepackage[T1]{fontenc}

\newcommand{\unit}[1]{\,{\rm#1}}

\DeclareFloatingEnvironment[
    fileext=loa,
    listname=List of Algorithms,
    name=ALGORITHM,
    placement=tbhp,
]{algorithm}

\usepackage{listings}

\lstdefinestyle{mystyle}{
    backgroundcolor=\color{lightgray},
    commentstyle=\color{red},
    keywordstyle=\color{blue},
    numberstyle=\tiny\color{gray},
    numbers=left,
    numbersep=5pt,
    showspaces=false,
    showstringspaces=false,
    showtabs=false,
    tabsize=2,
    breaklines=true,
    breakatwhitespace=true,
    title=\lstname
}
\usepackage{tcolorbox}
\tcbuselibrary{minted,breakable,xparse,skins}

\usepackage{hyperref}
\hypersetup{
    colorlinks=true,
    linkcolor=blue,
    filecolor=magenta,      
    urlcolor=blue,
    citecolor=blue,
    pdftitle={Overleaf Example},
    pdfpagemode=FullScreen,
    }

\definecolor{bg}{gray}{0.95}
\DeclareTCBListing{mintedbox}{O{}m!O{}}{%
  breakable=true,
  listing engine=minted,
  listing only,
  minted language=#2,
  minted style=default,
  minted options={%
    linenos,
    gobble=0,
    breaklines=true,
    breakafter=,,
    fontsize=\small,
    numbersep=8pt,
    #1},
  boxsep=0pt,
  left skip=0pt,
  right skip=0pt,
  left=25pt,
  right=0pt,
  top=3pt,
  bottom=3pt,
  arc=5pt,
  leftrule=0pt,
  rightrule=0pt,
  bottomrule=2pt,
  toprule=2pt,
  colback=bg,
  colframe=orange!70,
  enhanced,
  overlay={%
    \begin{tcbclipinterior}
    \fill[orange!20!white] (frame.south west) rectangle ([xshift=20pt]frame.north west);
    \end{tcbclipinterior}},
  #3}

\begin{document}

\title{Virtual gates enabled by digital surrogate of quantum dot devices}

\author{Alexander Lidiak$^\dag$}
\affiliation{QuantrolOx Ltd, 8 King Edward Street, Oxford, OX1 4HL, United Kingdom} 

\author{Jacob Swain$^\dag$}
\affiliation{QuantrolOx Ltd, 8 King Edward Street, Oxford, OX1 4HL, United Kingdom}

\author{David L.~Craig}
\affiliation{Department of Material Science, University of Oxford, Parks Road, Oxford OX1 3PJ, United Kingdom}
\affiliation{QuantrolOx Ltd, 8 King Edward Street, Oxford, OX1 4HL, United Kingdom}

\author{Joseph Hickie}
\affiliation{Department of Engineering Science, University of Oxford, Parks Road, Oxford OX1 3PJ, United Kingdom}

\author{Yikai Yang}
\affiliation{Department of Engineering Science, University of Oxford, Parks Road, Oxford OX1 3PJ, United Kingdom}

\author{Federico Fedele}
\affiliation{Department of Engineering Science, University of Oxford, Parks Road, Oxford OX1 3PJ, United Kingdom}

\author{Jaime Saez-Mollejo}
\affiliation{Institute of Science and Technology Austria, Am Campus 1, 3400 Klosterneuburg, Austria}

\author{Andrea Ballabio}
\affiliation{L-NESS, Physics Department, Politecnico di Milano, via Anzani 42, 22100, Como, Italy}

\author{Daniel Chrastina}
\affiliation{L-NESS, Physics Department, Politecnico di Milano, via Anzani 42, 22100, Como, Italy}

\author{Giovanni Isella}
\affiliation{L-NESS, Physics Department, Politecnico di Milano, via Anzani 42, 22100, Como, Italy}

\author{Georgios Katsaros}
\affiliation{Institute of Science and Technology Austria, Am Campus 1, 3400 Klosterneuburg, Austria}

\author{Dominic T.~Lennon$^*$}
\affiliation{QuantrolOx Ltd, 8 King Edward Street, Oxford, OX1 4HL, United Kingdom}

\author{Vincent P.~Michal$^{**}$}
\affiliation{Department of Engineering Science, University of Oxford, Parks Road, Oxford OX1 3PJ, United Kingdom}

\author{Erik M.~Gauger}
\affiliation{SUPA, Institute of Photonics and Quantum Sciences, Heriot-Watt University, Edinburgh EH14 4AS, United Kingdom}

\author{Natalia Ares}

\affiliation{Department of Engineering Science, University of Oxford, Parks Road, Oxford OX1 3PJ, United Kingdom}
\date{\today}

\begin{abstract}

Advances in quantum technologies are often limited by slow device characterization, complex tuning requirements, and scalability challenges. Spin qubits in electrostatically defined quantum dots provide a promising platform but are not exempt from these limitations. Simulations enhance our understanding of such devices, and in many cases, rapid feedback between measurements and simulations can guide the development of optimal design and control strategies. Here, we introduce a modular, graph-based simulator that acts as a digital surrogate for a semiconductor quantum dot device, where computationally expensive processes are accelerated using deep learning. We demonstrate its potential by estimating crosstalk effects between gate electrodes and applying these estimates to construct virtual gates in a quantum dot device. We validate our approach through comparison with experiments on a double quantum dot defined in a Ge/SiGe heterostructure. We envision that this simulation framework will advance semiconductor-based quantum technologies by enabling more efficient design, characterization, and control of complex devices.

\end{abstract}

\maketitle

\def\thefootnote{$\dag$}\footnotetext{These authors contributed equally to this work and are listed in alphabetical order.}

\def\thefootnote{$*$}\footnotetext{dominic@quantrolox.com} 
\def\thefootnote{$**$}\footnotetext{vincent.michal@nbi.ku.dk}

\section{Introduction}

The development of scalable quantum technology requires accurate and efficient methods for fabrication, characterization, and control. Progress is often slow, as exploring potential improvements typically demands time-consuming and resource-intensive experimental work. In recent years, machine learning methods have been employed to accelerate advances in quantum technologies \cite{gebhart2023learning,Ares2021Scalability, alexeev2024, acampora2025,genois2021quantum, flurin2020using,ji2023recent}, and several approaches have leveraged the additional insights gained from simulations in the context of physics-aware machine learning \cite{craig2024bridging,wright2022deep,berritta2024}. In such approaches, simulations must be sufficiently fast to provide timely feedback -- for example, to guide device design or to support characterization and control algorithms.

Electrostatically defined quantum dots in semiconductors are a well-established platform where several characterization and control approaches have been developed. These devices have a wide range of applications including quantum computation \cite{Vandersypen2017,Scappucci2021,Burkard2023, steinacker2024300, takeda2024rapid} and quantum simulation \cite{Hensgens2017,Dehollain2020,Kiczynski2022}. In particular, encoding qubits in the spin states of electrons or holes confined in electrostatically-defined quantum dots \cite{Loss1998} promises a scalable platform for quantum computing with favorable operating temperatures \cite{Vandersypen2017,Huang2024, van2024advanced}. These devices are also used in foundational studies, such as probing quantum transport phenomena \cite{hanson2007spins,van2002electron,hartke2018microwave,hofmann2020phonon}, and exploring thermodynamics at the nanoscale \cite{josefsson2018quantum, majidi2022quantum, aggarwal2025rapid, wadhia2025entropic}. Quantum dots are defined and controlled by voltages applied to gate electrodes, which confine single spins and enable fast qubit manipulation~\cite{Vandersypen2017,Philips2022,Noiri2022}. Finding appropriate gate voltage configurations that reliably define quantum dot potentials and enable precise qubit control is challenging, as they are sensitive to cross-talk between gate electrodes and to electrostatic disorder arising from fabrication variation and unavoidable imperfections in the semiconductor~\cite{croot2019gate,klos2018calculation,craig2024bridging}.
Recent efforts have developed automated tuning algorithms using machine learning to reduce reliance on extensive human expertise and to accelerate tuning times \cite{Moon2020,severin2021cross,van2022all,Zwolak2023Rev,schuff2024fully, che2024fast, RaoMavis2025}. As
the number of quantum dots increases, the higher density of gate electrodes required for operation leads to stronger capacitive crosstalk due to their close proximity to one another and to the dots. This increased electrostatic coupling hinders independent control of quantum dots, making the tuning process more experimentally demanding and time-consuming ~\cite{Hensgens2017,vanDiepen2018,Volk2019,Fedele2021}.
In existing quantum dot simulators~\cite{vanStraaten2024QArray,Greplova2025,Krzywda2025} and tuning algorithms~\cite{RaoMavis2025,hickie2023automated}, estimates of crosstalk effects are based on a minimal model of constant capacitors. As a result, mitigation strategies remain effective only within a narrow range of gate voltage parameters.

In this article, we present a digital surrogate for an electrostatically defined quantum dot device and demonstrate its effectiveness in addressing effects such as crosstalk. We introduce a modular approach which structures the digital surrogate as a directed acyclic graph where each node performs a computation relevant to the complete simulation and edges pass information. Nodes involving slow computation are replaced with fast deep-learning approximations to facilitate practical simulation times. Although our graph-based structure is general, we apply our simulator to devices defined in Ge/SiGe heterostructures \cite{jirovec2021singlet}, and compare simulation results with experiments. In particular, we use our simulator to evaluate methods for computing so-called virtual gates, i.e. linear combinations of gate voltages specifically chosen to compensate for crosstalk effects. 
Although gate virtualization approaches have been demonstrated~\cite{Hensgens2017, vanDiepen2018, Volk2019, Hsiao2020, RaoMavis2025}, our method constructs virtual gates solely based on electrostatic simulations within a unified framework and provides deeper insight into the electrostatic environment of the devices, which is difficult to probe in experiments. We validate our method by identifying a virtualization procedure in simulation, and by confirming its success both in simulation and on the physical device. We anticipate that our graph-based simulator architecture will support future device design processes and serve as a testbed for automated characterization and control algorithms by enabling efficient computation of relevant device properties.

\begin{figure*}[t]
    \includegraphics[width=\textwidth]{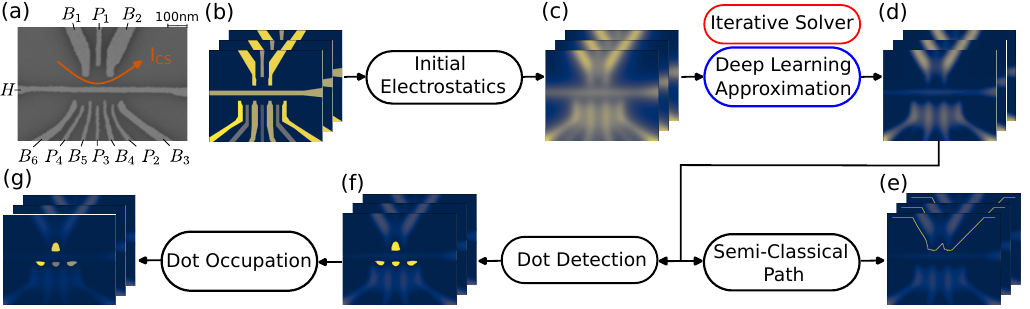}
    \caption{(a) SEM image (880x750 nm) of a device architecture similar to that modeled with the simulator; referred to as the \textit{experimental device} (its design is shown in Fig.~\ref{fig:experimental_device_gate_design}). (b-g) Sub-results produced by the graph-based simulator. (b) Applied gate voltages. (c) Initial electrostatic potential arising from gate voltages, surface potentials, and random disorder. (d) Self-consistent potential obtained through iterative solutions of the electric field generated by the 2DHG charge density. The color intensity (yellow) indicates potential magnitude. The iterative solver (red) can be replaced by a trained deep learning model (blue) to accelerate computations. With the self-consistent potential, we compute subsequent quantities, including (e) the charge transport semi-classical trajectory (yellow path), (f) quantum dot formation (yellow blobs), and (g) dot charge occupancy (yellow intensity).}
    \label{fig:device_graph}
\end{figure*}

\section{Simulator}\label{sec:graph_simulator}

Our simulator is built on a flexible graph structure (Fig.~\ref{fig:device_graph}), where each node can be independently configured to perform specific computational tasks using a range of methods, including exact solvers, iterative algorithms, and machine learning-accelerated approximations. Starting from the gate electrode design, this graph-based architecture integrates nodes that compute the electrostatic environment, including confining potentials and charge densities, with nodes that simulate device behavior, enabling the prediction of optimal parameters for experimental operation. This modular and extensible framework allows realistic device physics to directly inform control strategies, enhancing both the accuracy and scalability of tuning algorithms.

\subsection{Graph-Based Simulation}
\label{subsec:graph-based-simulation}
In Fig.~\ref{fig:device_graph}, we show how the end-to-end simulation process is decomposed into individual nodes, each computing a specific sub-result. The variables defining the physical properties of the device are contained in the system configuration, and each node may receive additional parameters required for subsequent computational tasks (Appendix~\ref{app:graph-config}).

Figures~\ref{fig:device_graph}(d-e) highlight how the modular structure of the simulator supports the inclusion of additional computational models as separate branches in the directed graph. In this example, after computing the self-consistent electrostatic potential (Fig.~\ref{fig:device_graph}d), we add nodes to predict tunnel rates based on Landauer's scattering formalism and its semi-classical approximation, as well as to estimate tunnel rates from the dots to nearby reservoirs (Fig.~\ref{fig:device_graph}e), which is particularly relevant for readout protocols~\cite{elzerman2004single,Schuff2023identifyingpauli} (see Appendices \ref{app:self_consistent_pot}-\ref{app:tunnel_semiclassics} for details). This structure also facilitates efficient evaluation of outputs by computing only the necessary prerequisite quantities or by starting from precomputed data, rather than evaluating all nodes in the graph. Individual nodes may also be accelerated by parallelization.

\subsection{Deep learning acceleration}
\label{subsec:deep-learning-acceleration}
Computationally intensive nodes that involve iterative or slow physical predictions can be replaced by deep learning models. This is shown schematically between Figs.~\ref{fig:device_graph}(c) and (d) for a single node: a time-consuming prediction obtained with an iterative solver is replaced by a deep-learning model that reproduces the quantity with significantly reduced runtime. Once trained, the deep-learning model approximates the physical computation with orders-of-magnitude speedup over the original node. 

A bottleneck in our simulation is the iterative calculation of the self-consistent electrostatic potential required to estimate the charge density in the two-dimensional hole gas (2DHG), so its acceleration is key to practical simulation times. To achieve this, we utilize a fully convolutional neural network (CNN) with residual blocks like in the ResNet model \cite{he2016deep}. The CNN takes gate voltages, surface potentials, and disorder potentials as input and approximates the self-consistent potential~\cite{craig2024bridging}. 
    
To train the CNN, realistic electrostatic potential datasets were generated using a self-consistent iterative solver. The datasets include different device designs to promote model generalization. Random gate architectures were created by sampling individual electrode designs from a batch, with each design randomly oriented and scaled. The voltages of the electrodes were sampled from a uniform distribution that spanned typical operating ranges. Uniform sampling was also applied to surface potential offsets, donor depth, charge density, and defect density. To bias the dataset toward conditions that exhibit quantum dots, results from the iterative solver were passed to the simulator node that checks for their presence (Fig.~\ref{fig:device_graph}f); if no dots were detected, the input parameters were resampled once. 

Model hyperparameters -- namely the number of residual blocks, the number of channels per block, and the convolution kernel size -- were chosen by sampling various hyperparameter sets, training a model for each set, and selecting the model that achieved the best performance as measured by evaluation metrics (see Appendix~\ref{app:graph_and_results} for details and examples). 

Using the deep-learning model to compute the electrostatic potential provides an acceleration of $O(10^3)$ over the equivalent physical solver. This node is critical, as it enables an overall acceleration of $O(10^2)$ when evaluating the entire graph.

\section{Results}
Throughout the results section, we use our simulator as a digital surrogate for a quantum dot device in a Ge/SiGe heterostructure. Figure~\ref{fig:device_graph}(a) shows an electron micrograph of a similar device architecture used as a reference to qualitatively support the results of the digital surrogate simulation. The architecture of the device is designed to support the formation of up to four quantum dots. Ohmic contacts provide two transport channels: one through a linear array of three quantum dots, controlled by plunger gates $P_{2-4}$ and barrier gates $B_{3-6}$, and the other through a separate quantum dot, used as a charge sensor, and controlled by gate voltages $B_1$, $P_1$, and $B_2$. The sensor dot is separated from the array by the horizontal gate $H$. A bias voltage $V_{\text{bias}}$ applied to Ohmic contacts drives a current $I_\mathrm{CS}$ through the charge sensor dot to probe the charge states of the dots in the array.

\subsection{Stability Diagrams}
Operation of electrostatically defined quantum dots as qubits requires estimating their charge occupations, often visualized in a charge stability diagram \cite{vanderWiel2002}. Figure~\ref{fig:virtgates_in_exp}(c) shows a stability diagram obtained from charge sensing measurements of our experimental device configured to exhibit a double quantum dot. Sets of parallel diagonal lines appear in the sensor dot current, corresponding to charge transitions in each dot.

To demonstrate that the simulator can reproduce qualitatively similar data, we configure it to define two quantum dots and a sensor dot, with gate voltages matching those used in the experiment. The resulting simulated stability diagram is shown in Fig.~\ref{fig:virtgates_in_exp}(b). The simulation captures key features at the same gate voltages, such as the cross-capacitance between dots and gates as well as inter-dot capacitances. We observe some discrepancies between the simulation and the experiment. First, the capacitance between the charge sensor dot and gates $P_3$ and $P_4$ differs, with the simulation showing weaker coupling to $P_3$ than the experiment. This may arise from unaccounted differences between the SEM image used to construct the simulation and the measured device, or from discrepancies in the generated potential due to disorder effects \cite{craig2024bridging}. Furthermore, the simulation is carried out in the limit of negligible inter-dot tunnel coupling. This coupling can be estimated by supplementing the simulator with semiclassical methods \cite{DasSarma2011} (Appendix~\ref{app:tunnel_semiclassics}).

\subsection{Virtual gates}\label{sec:virtual_gates}

\begin{figure*}[hbt!]
    \centering
    \includegraphics[width=\textwidth]{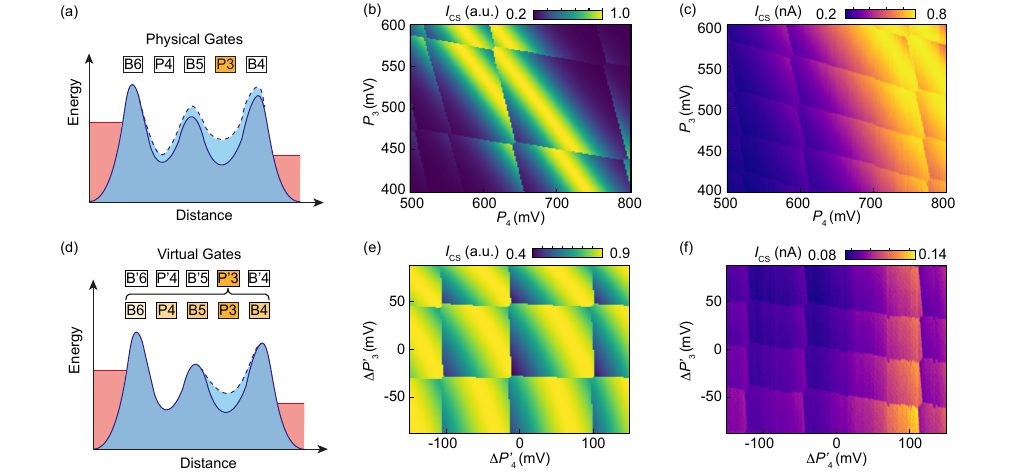}
    \caption{
    (a) Schematic of a double quantum dot confinement potential controlled by five gate electrodes. The potential energy is indicated in blue and Fermi reservoirs in red. The shade of yellow in the gate electrode labels indicates the relative voltage strength. As an illustration of crosstalk effects, the lighter blue potential exemplifies how varying gate electrode voltage $P_3$ also shifts the electrochemical potential of the nearest dot and the interdot barrier. (d) Varying the virtual gate $P'_3$ produces a more localized effect. (b–c, e–f) Charge-sensor current for the device configured as a double quantum dot defined with plunger gates $P_3$ and $P_4$. Experimental data are shown in (c,f), and simulated data from the surrogate in (b,e), taken in the same gate voltage region.
    The stability diagrams are generated by sweeping $P_3$ and $P_4$ (b–c) or by applying the virtual gates $P'_3$ and $P'_4$ estimated with the simulator (e–f). In (e–f), the charge-transition lines are closer to orthogonal than in (b–c), indicating that the virtual gates enable independent control of the dot electrochemical potentials. The virtual gate voltages estimated with the simulator also work well when applied to the real device, leaving only minimal residual crosstalk. Compared with the real device, the surrogate displays a higher addition energy, evidenced by fewer charge transitions, particularly for the dot adjacent to $P_3$. Possible sources of discrepancy include the breakdown of semiclassical approximations at low charge occupation, spin-filling effects, and presence of trapped carriers. In both the surrogate (e) and experimental device (f), the virtual gate sweeps include the charge sensor plunger, which maintains high readout contrast in the sensing signal.
    }
    \label{fig:virtgates_in_exp}
\end{figure*}

Cross-capacitance is evident from the diagonal charge transition lines in both Figs.~\ref{fig:virtgates_in_exp}(b) and (c). If we focus on small regions of voltage space, the gate cross-capacitance is approximately linear and the effect of gate voltages on the tunnel coupling is well described by an exponential. To parametrize these dependencies, we evaluate the derivative of the dot potential wells and tunnel barriers with respect to changes in each gate's voltage. The full set of derivatives forms a crosstalk matrix that relates the real gate space $\boldsymbol{G}=\{P_i,B_i\}$ to the \textit{virtual gate} space $\boldsymbol{G}'=\{P_i',B_i'\}$. The matrix for two lateral dots is:

\begin{equation}
\begin{bmatrix}
    P_1 \\
    P_2 \\
    B_3 \\
    B_4 \\
    B_5 \\
\end{bmatrix}
=
\begin{bmatrix}
    1 && \alpha_{12} && \alpha_{13} && \alpha_{14} && \alpha_{15} \\
    \alpha_{21} && 1 && \alpha_{23} && \alpha_{24} && \alpha_{25} \\
    \beta_{31} && \beta_{32} && 1 && \beta_{34} && \beta_{35} \\
    \beta_{41} && \beta_{42} && \beta_{43} && 1 && \beta_{45} \\
    \beta_{51} && \beta_{52} && \beta_{53} && \beta_{54} && 1 \\
\end{bmatrix}
\begin{bmatrix}
    P_1' \\
    P_2' \\
    B_3' \\
    B_4' \\
    B_5' \\
\end{bmatrix},
\label{eq:virtual_gates}
\end{equation}
where $\alpha_{ij} = \frac{\partial \mu_i}{\partial G_j}/\frac{\partial \mu_i}{\partial P_i}$, ($\beta_{ij} = \frac{\partial \tau_i}{\partial G_j}/\frac{\partial \tau_i}{\partial B_i}$) are the normalized derivatives of the $i^\mathrm{th}$ dot electrochemical potential $\mu_i$ (the $i^\mathrm{th}$ tunnel barrier $\tau_i$) with respect to changes in gate voltage $G_j$. The inverse of this crosstalk matrix defines the linear combination of real gates that form the virtual plunger ($P_i'$) and virtual barrier gates ($B_i'$), enabling orthogonal control of the dot electrochemical potentials and barriers, respectively. Virtual gates remain most effective when operating in gate voltage regions close to where the crosstalk matrix was evaluated.

Numerically determining both the $\alpha_{ij}$ and $\beta_{ij}$ terms with the simulator is straightforward. In particular $\alpha_{ij}$ can be approximated as the numerical derivative of  $\mu_{i}$ which is a direct simulator output. The tunnel coupling $\tau_{i}$ can be approximated as the exponential of the semiclassical action (i.e. the solution of the semiclassical Schr{\"o}dinger equation for the optimal path through barrier $i$, denoted $\mathrm{minpath}[B_i]$ \cite{LLbook, Banks1973}). In this case, the matrix element for tunneling through barrier $i$ takes the form $\tau_{i} = \tau_0\exp(\phi_i)$, with $\phi_i = -\int{\,ds\sqrt{2 m (U_i(s) - \varepsilon)}}/\hbar$ for $s \in \mathrm{minpath}[B_i]$. Using the equivalence: $ \beta_{ij} = \frac{\partial \tau_i}{\partial G_j}/\frac{\partial \tau_i}{\partial B_i} =   \frac{\partial \phi_i}{\partial G_j}/\frac{\partial \phi_i}{\partial B_i}$, $\beta_{ij}$ can be directly derived from the self-consistent potential landscape along the estimated charge transport path through the barrier.
In experimental settings, a common strategy to estimate the $\alpha$ terms (lever arms) is to monitor the position of a Coulomb peak as a function of two different gate voltages. The slope of the observed Coulomb peak in this two-dimensional gate sweep provides a local linear approximation of the corresponding $\alpha$ term, capturing the capacitive coupling between the two gates \cite{Volk2019, Hensgens2017, vanDiepen2018, Zwolak2023Rev}.

Once $\alpha_{ij}$ is determined for each dot-gate pair, the pseudo-inverse of the $\alpha$ submatrix forms the virtual plunger gates $P_i'$. The $\beta_i$ terms can be characterized experimentally by fitting the tunnel rates as a function of the charge transition positions \cite{Hsiao2020}. The pseudo-inverse of the derived $\beta$ submatrix then defines virtual barrier gates that control the quantum dot tunnel barriers while mitigating the cross capacitance of the dot electrochemical potentials. 

Because each submatrix represents an underdetermined system, separate characterization of $\alpha$ and $\beta$ provides only partial orthogonalization. Full orthogonalization, enabling independent control of both the dot electrochemical potentials and the barrier tunnel couplings, requires estimating and inverting the complete crosstalk matrix in Eq.~\eqref{eq:virtual_gates}. In practice, experimentally determining all the gradients needed to construct both the $\alpha$ and $\beta$ submatrices is often impractical and time-consuming. 
In contrast, in the surrogate device, the potential landscape is directly observable, allowing the full crosstalk matrix to be generated straightforwardly by varying each gate voltage $G_i$ and recording the numerical gradients $\frac{\partial \mu_j}{\partial G_i}$ for each dot and $\frac{\partial \phi_j}{\partial G_i}$ for each barrier. Crucially, the surrogate not only provides the full crosstalk matrix but does so efficiently, as all numerical derivatives ($\Delta\mu_i$ and $\Delta\tau_i$) follow from a single gate voltage sweep, eliminating the need to locate charge transitions. Importantly, over large gate voltage sweeps, this efficiency allows us to rapidly re-estimate the full-crosstalk matrix when necessary to maintain the validity of the local approximation. 

\begin{figure}[ht!]
    \includegraphics[width=0.49\textwidth]{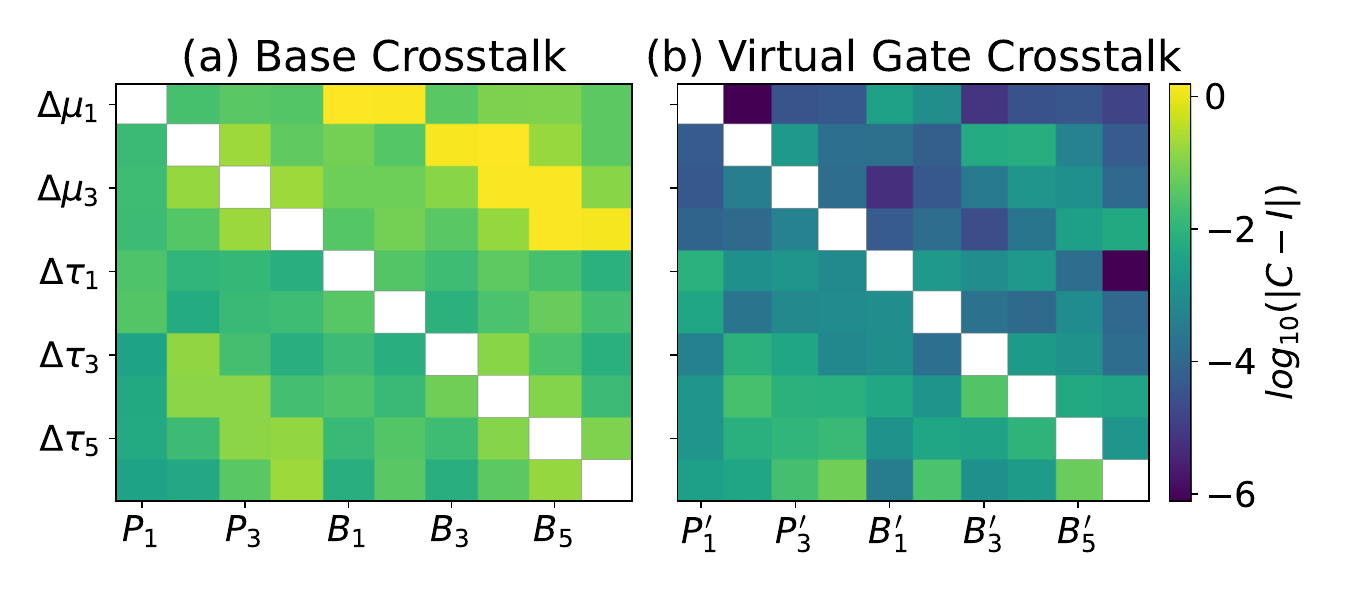}
    \caption{Observed changes in dot electrochemical potentials ($\Delta \mu_i$) and barrier tunnel couplings ($\Delta \tau_i$) normalized by the diagonal form the crosstalk matrix $\boldsymbol{C}$ as in Eq.~\eqref{eq:virtual_gates}. Each non-white pixel represents the quantity $\log_{10}{|{\boldsymbol{C}-\boldsymbol{I}}|}$ calculated from the crosstalk matrix $\boldsymbol{C}$, estimated from the surrogate while sweeping non virtualized (a) and virtualized gate voltages (b). The virtual gates used in (b) are derived by inverting the crosstalk matrix in (a). (b) Highlights how the surrogate compensation parameters effectively mitigate cross-talk in controlling dot electrochemical potentials and tunnel barriers.
    }
    \label{fig:crosstalk_matrices}
\end{figure}

We use this approach to generate a simulated crosstalk matrix and invert it to generate virtual gates. We apply these simulated virtual gates to both the surrogate device and the experimental device to generate the stability diagrams in Fig.~\ref{fig:virtgates_in_exp}(e) and Fig.~\ref{fig:virtgates_in_exp}(f), respectively. The simulated virtual gates significantly reduce crosstalk for both the surrogate device and the experimental device. The inclusion of the charge sensor plunger in the gate virtualization also allows us to maintain a high readout contrast, as the charge sensor electrochemical potential remains approximately constant as the double quantum dot is swept over multiple charge transitions. Vitally, the virtualization parameters predicted by the surrogate lead to a substantial reduction in crosstalk in the experimental device (Fig.~\ref{fig:crosstalk_matrices} and Tab.~\ref{table:crosstalk_mats_diff_from_I}), even though the compensation observed in the experimental device does not exactly match that of the surrogate. The residual discrepancies may arise from deviations between the fabricated gate geometry and the design (Fig.~\ref{fig:experimental_device_gate_design}), unmodeled imperfections in the device - such as localized pools of charge carriers - and limitations of the semiclassical approximation at low charge occupation.

To evaluate the efficacy of the virtualization coefficients, we use the surrogate to estimate the crosstalk matrix parameters before and after applying the gate compensation. To illustrate this comparison, in Fig.~\ref{fig:crosstalk_matrices}(a-b) we show the quantity $\log_{10}{|{\boldsymbol{C} - \boldsymbol{I}}|}$, where $\boldsymbol{I}$ is the identity matrix, and $\boldsymbol{C}$ is the crosstalk matrix estimated using non-virtualized (panel a) and virtualized (panel b) gate voltages. In the ideal case of perfectly compensated control $\boldsymbol{C} = \boldsymbol{I}$,  smaller coefficients indicate better crosstalk mitigation achieved by gate virtualization. The reduced off-diagonal elements presented in (Fig.~\ref{fig:crosstalk_matrices}b) in comparison to those extracted from the uncompensated case (Fig.~\ref{fig:crosstalk_matrices}a), highlight the crosstalk mitigation achieved by our gate virtualization. Our results, estimated from the surrogate, are summarized in Table~\ref{table:crosstalk_mats_diff_from_I} where we present the normalized average of the block off-diagonal elements of the crosstalk matrix without gate voltage compensation (Base), with compensation parameters estimated from the submatrix approach (Submatrix), and from the inversion of the full crosstalk matrix (Full Matrix). Overall, our virtualization method achieves an average improvement of almost two orders of magnitude with the full matrix approach.

\begin{table}[ht]
\setlength{\tabcolsep}{10pt}
\centering
\begin{tabular}{|c|c|c|c|}
\hline
 \textbf{Virtual Gate} & \multicolumn{3}{c|}{\textbf{Gate Space}} \\
\cline{2-4}
\textbf{Type} & Complete & $\mu$ & $\tau$ \\
\hline
Base & 0.13 & 0.096 & 0.037 \\
Submatrix & 1.0e-2 & 2.8e-3 & 7.2e-3  \\
Full Matrix & 6.0e-3 & 2.7e-3 & 3.4e-3 \\
\hline
\end{tabular}
\caption{Comparing virtual gate estimation methods in compensating crosstalk effects on the dot potential ($\mu$) and barrier tunnel coupling ($\tau$) subspaces.
Each value in the table represents the average of the estimated off-diagonal crosstalk matrix parameters, each normalised by its respective diagonal element.
The leftmost column lists the virtual gate type; `Base' shows results from the crosstalk matrix estimated from the surrogate without prior compensation, `Submatrix' shows the results from the crosstalk matrix obtained while sweeping virtual plungers and barriers, estimated separately from the $\alpha$, and $\beta$ submatrices~\cite {Volk2019, Hsiao2020}. `Full matrix' shows the results obtained after virtualising the gate voltages from the full crosstalk matrix inverse as in Eq.~\eqref{eq:virtual_gates}.
}
\label{table:crosstalk_mats_diff_from_I}
\end{table}

\section{Conclusion}

We have introduced a machine-learning-accelerated graph-based simulator of electrostatically defined semiconductor quantum dot devices.

We have applied our simulator as a digital surrogate of a Ge/SiGe device for which we calculate virtual gates. In the stability diagrams of both the surrogate and experiment, the simulated virtual gates significantly reduced crosstalk. Unlike existing approaches, which typically rely on incomplete crosstalk matrices \cite{Volk2019,Hsiao2020, vanDiepen2018} and involve multiple stages of analysis \cite{RaoMavis2025}, our method leverages electrostatic simulations to characterize the entire crosstalk matrix directly.

The modular structure of our simulator enables future extensions through the addition of nodes to estimate new device properties or the replacement of existing nodes with refined, updated, or accelerated models. Such extensions could incorporate spin degrees of freedom for quantum computation \cite{Xue2022,Mills2022}, or optimize spin dynamics under time-dependent voltages, for instance in spin shuttling experiments \cite{vanriggelendoelman2023coherent, wang2024operating}. In these regimes, versatile simulations are vital, as optimal operation points require even finer tuning of gate voltages \cite{schuff2024fully, dumoulin2024silicon}, and optimal control of spins requires precisely tuned pulses \cite{Michal2023,wang2024operating}. 

Our work shows that efficient machine learning methods can enhance the combined use of simulation and experiment to characterize quantum systems~\cite{gebhart2023learning}. Relevant applications include characterizing disorder in semiconductor devices \cite{craig2024bridging}, inferring from experiment the parameters of a Hamiltonian \cite{wang2017experimental} or Lindbladian \cite{genois2021quantum,craig2024differentiable} representation of the system, and learning non-Markovian dynamics \cite{krastanov2020unboxing}. Overall, we anticipate that simulations based on physical models and deep learning will play an increasingly important role in advancing the design and control of electrostatically-defined quantum dot devices, further enhancing their position as a platform for probing quantum phenomena.

\begin{acknowledgments}
We acknowledge funding from Innovate UK Grant Number 10031865. We also acknowledge Juha Seppa and Frederick Faulkner for their support throughout the development of this work, and Frederico Martins for insightful discussions. G.K. acknowledges support from HORIZON-RIA (project no. 101069515) and the FWF Project (DOIs: 10.55776/F86).
\end{acknowledgments}

\section*{Author Contributions}

A.L. and J.S. developed the model in collaboration with D.L.C. and V.P.M. with support from D.T.L. and E.M.G. Experiments were performed by J.H. and Y.Y. with assistance from D.L.C.,V.P.M and F.F. The sample was fabricated by J.S.M. and G.K. The project was conceived by D.T.L. and N.A. The manuscript was written by A.L., J.S., D.L.C., V.P.M, and F.F. with input and discussion from all authors.

\clearpage

\twocolumngrid

\bibliographystyle{apsrev4-1}
\bibliography{bibliography_file}

\clearpage
\appendix
\renewcommand{\theequation}{\thesection\arabic{equation}}

\section{Graph-Based Simulator Configuration}
\label{app:graph-config}

\begin{figure}[ht!]
    \centering
    \includegraphics[width=0.49\textwidth]{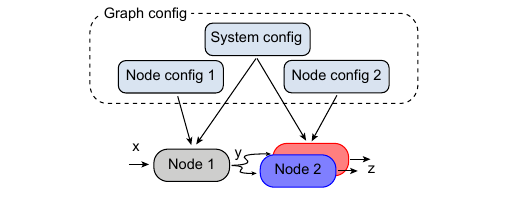}
    \caption{Schematic showing how configuration information (such as the device schematic in Fig.~\ref{fig:experimental_device_gate_design}) is distributed to the graph. The system configuration is needed for each node, and node configurations are specific to a given node.}
    \label{fig:graph_config}
\end{figure}

As discussed in the main text, the graph based simulator is initialized with information pertaining to the device under consideration and the desired outputs from the graph. The graph configuration comprises of a system configuration and node configurations. The system configuration holds data pertinent to all nodes, and a node configuration contains information relevant to a particular computation and its connectivity within the graph. For example, details of device geometry and material properties would be contained in the system configuration, while lattice dimensions for a numerical solver would be contained in a node configuration. A schematic of how this information is distributed is shown in Fig.~\ref{fig:graph_config}. To further improve efficiency, a recursive graph traversal algorithm ensures that only the necessary nodes are computed when inputting suitable pre-computed data.

\begin{figure}[ht!]
    \includegraphics[width=0.33\textwidth]{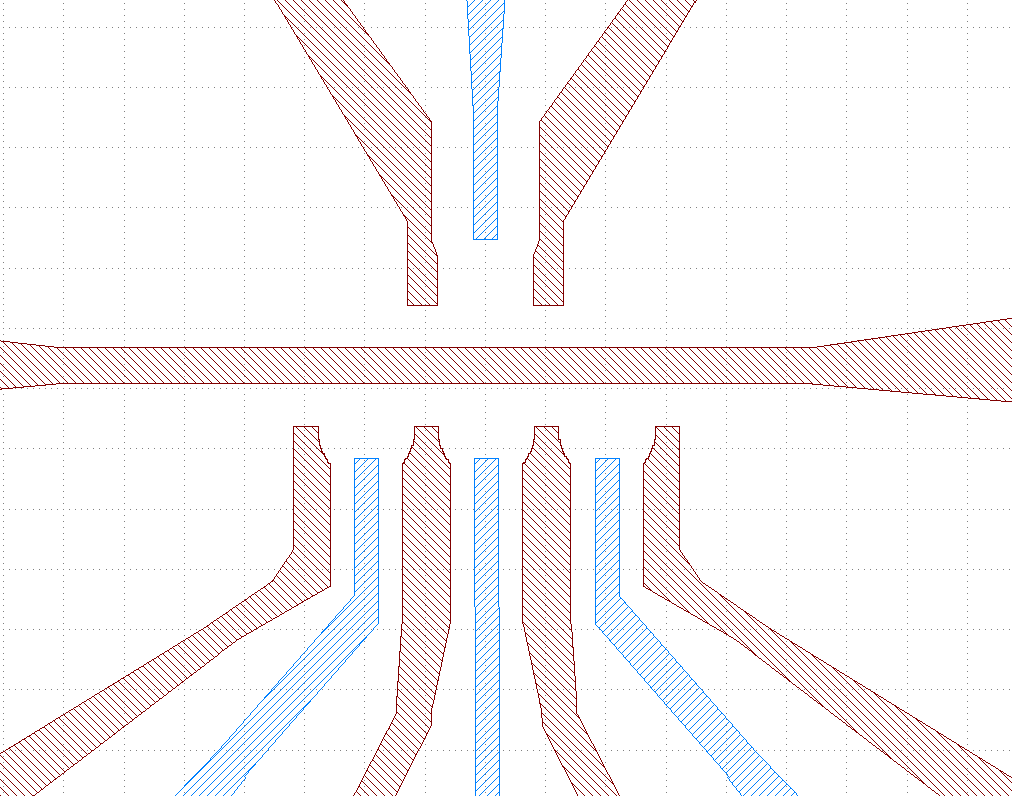}
    \caption{
        Gate electrode design of the device used to configure the surrogate device gates.
    }
    \label{fig:experimental_device_gate_design}
\end{figure}

\section{Self-consistent electrostatic potential model} \label{app:self_consistent_pot}

The charge configurations in the quantum dots are computed with the Thomas-Fermi model for the self-consistent electrostatic potential \cite{Stopa1996,IhnBook} (see \cite{craig2024bridging} for a detailed formulation of the model and its deep learning integration). We furthermore solve the two-dimensional finite-difference Schr\"odinger equation in the self-consistent potential using existing (open-source) software \cite{Groth2014}. In Appendix~\ref{app:Landauer} we discuss the Landauer formulation for computing current through a single quantum dot, and relate it to the self-consistent potential model. Then in Appendix~\ref{app:tunnel_semiclassics} we compute the tunnel rates relevant to quantum transport equations, benchmark their semi-classical expressions with numerics (Fig.~\ref{fig:semiclassical_tunnel}), and compute their values for a realistic device design. 

\section{Quantum dot transport from the Landauer viewpoint}
\label{app:Landauer}

We aim to numerically estimate the tunnel rates from the current through a single quantum dot using the Landauer formula for the average current \cite{Datta_1995,NazarovBook}, with a single spin channel:
\begin{equation}\label{eq:Landauer}
    I = \frac{e}{\hbar}\int_{-\infty}^{+\infty}d\varepsilon\,\mathcal{T}(\varepsilon)[f(\varepsilon-\mu_L)-f(\varepsilon-\mu_R)],
\end{equation}
where $\mathcal{T}(\varepsilon)$ is the transmission probability across the single quantum dot, and $f(\varepsilon-\mu) = \big[\exp\big(\frac{\varepsilon-\mu}{k_B T}\big)+1\big]^{-1}$ the Fermi-Dirac distribution characterized by the temperature $T$ and the electrochemical potential $\mu$. Here $k_B$ is the Boltzmann constant and $\hbar$ is the reduced Planck constant. When transport occurs through a single level in the dot the transmission function has the Breit-Wigner form for resonant tunneling \cite{NazarovBook}
\begin{equation}\label{eq:Breit-Wigner}
    \mathcal{T}(\varepsilon)=\frac{\Gamma_L\Gamma_R}{(\varepsilon-\varepsilon_0)^2/\hbar^2+(\Gamma_L+\Gamma_R)^2/4},
\end{equation}
where $\varepsilon_0$ is the energy of the resonance in the quantum dot. 

Following from Eq. (\ref{eq:Landauer})) and Eq.(\ref{eq:Breit-Wigner}), the linear conductance at high temperature ($4k_BT\gg\hbar[\Gamma_L+\Gamma_R]$) evaluates to 
\begin{equation}
    G = G_Q\frac{\hbar\Gamma_L\Gamma_R/(\Gamma_L+\Gamma_R)}{4k_BT\cosh^2\left(\frac{\overline{\mu}-\varepsilon_0}{2k_BT}\right)},
\end{equation}
with $G_Q=e^2/h$ the quantum of conductance for a single spin channel, and $\overline{\mu}=(\mu_L+\mu_R)/2$. The conductance peak is sharpest  at low temperature ($4k_BT\lesssim\hbar(\Gamma_L+\Gamma_R)$) where,
\begin{equation}
    G=G_Q\mathcal{T}(\overline{\mu}).
    \label{eq:low-temp-resonance}
\end{equation}

In the Coulomb blockade regime the conductance peaks are separated by the gate voltages $\Delta V_g(N)=[\mu(N+1)-\mu(N)]/e\alpha_g$, where $\mu(N)=E(N)-E(N-1)$ is the electrochemical potential of the quantum dot containing $N$ charges, and $E(N)$ the energy in the ground state configuration \cite{hanson2007spins}. The lever arm of the gate electrode, $\alpha_g$, converts voltage to dot energy levels. Conductance peaks in the Coulomb blockade regime are well described by the equations above when the broadening is dominated by the coupling to the leads at low temperatures. In other regimes more general expressions for the conductance may be considered \cite{Beenakker1991, Meir1991, Datta_1995}. The current through the sensor quantum dot also acts as a probe of the tunnel coupling between neighbouring dots \cite{DiCarlo2004, Hensgens2017}. 

\section{Tunneling in the semi-classical case}
\label{app:tunnel_semiclassics}

In a semi-classical description of quantum mechanics \cite{LLbook}, the Hamiltonian matrix element for tunneling through a barrier of a double well potential is
\begin{equation}\label{eq:tunnel}
    \tau = a\hbar\nu\exp\Big(-\int_{s_1}^{s_2}ds\,\kappa(s)\Big),
\end{equation}
with $\kappa(s)=\sqrt{2m(U(s)-\varepsilon)}/\hbar$ and the integral evaluated along the optimal semi-classical path through the classically forbidden region of the potential energy barrier \cite{Banks1973}. In addition, $\nu=\left(\int ds\sqrt{\frac{2m}{\varepsilon-U(s)}}\right)^{-1}$ is the typical classical frequency of motion along a single dimension of the dot, with the integration performed over the classically allowed segment of the trajectory. In Eq. (\ref{eq:tunnel}) we introduce the numerical factor, $a$, to account for the 2D geometry of the dots and take it as a parameter for fitting with numerical data. The tunneling matrix element is expected to primarily depend on the energy barrier profile rather than on the precise dot structure. The tunnel rate, $\Gamma$, between the dot and a nearby reservoir is proportional to the semi-classical transmission probability through the barrier $\mathcal{T}_\text{barrier}=\exp\left(-2\int_{s_1}^{s_2} ds\,\kappa(s)\right)$ \cite{LLbook, NazarovBook}, and is evaluated with Fermi's golden rule as 
\begin{equation}\label{eq:tunnel_rate}
\Gamma=\frac{\pi m s_0^2\tau^2}{\hbar^3},
\end{equation}
where $s_0$ is the semi-classical radius of the dot, and assuming that the reservoir is well represented by a region of area $\pi s_0^2$, in close proximity to the quantum dot.

\begin{figure}[ht!]
    \centering
    \includegraphics[width=0.45\textwidth]{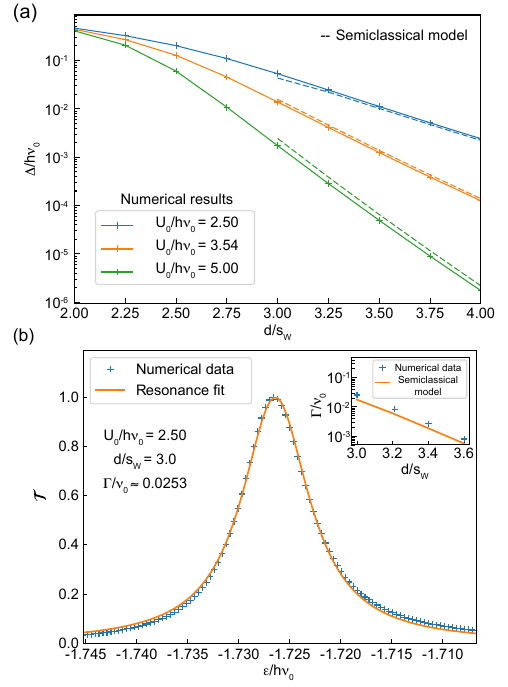}
    \caption{Numeric and semi-classical results for quantum tunneling in 2D. a) Energy separation, $\Delta$, between the symmetric and anti-symmetric states of the double quantum dot for different values of $U_0/h\nu_0$, compared with the semi-classical formula Eq.~\eqref{eq:tunnel} with $a=0.25$. The inter-dot tunnel energy, $\tau=\Delta/2$, is expressed in units of the oscillator level separation $h\nu_0$, with $h=2\pi\hbar$ the Planck constant, and the inter-dot distance in units of the potential well radius $s_w$. b) Resonant tunneling probability through a single quantum dot symmetrically positioned between two infinite reservoirs, numerically calculated with the Landauer formalism presented in Appendix~\ref{app:Landauer}. The tunnel rate $\Gamma$ is extracted as the half-width at half-maximum of the transmission probability resonance. Here the parameters are $U_0/h\nu_0=2.5$ and $d/s_w=3$. The inset displays the dependence of the tunneling rate on the separation between the dot centers, and compares the numerical data and semi-classical predictions.}
    \label{fig:semiclassical_tunnel}
\end{figure}

We numerically benchmark the semi-classical tunneling model in two dimensions using a simple model for the quantum dots and the reservoirs. We compute the quantum transport properties using \texttt{KWANT} \cite{Groth2014}. The system of $N$ quantum dots is modeled using the potential energy $U(\mathbf{r})=-U_0\sum_{i=1}^{N}\exp\big(\frac{-|\mathbf{r}-\mathbf{r}_i|^2}{2s_w^2}\big)$; so an individual potential well is characterized by the oscillator frequency $\nu_0=\tfrac{1}{2\pi s_w}\sqrt{\tfrac{U_0}{m}}$, typically from tens of $\unit{GHz}$ to the ${\rm THz}$ range.
The tunneling matrix element is computed from the energy splitting, $\Delta$, of the symmetric and anti-symmetric states of the double quantum dot system as $\tau=\Delta/2$. We show this quantity as a function of the inter-dot distance $d=|\mathbf{r}_1-\mathbf{r}_2|$ in Fig.~\ref{fig:semiclassical_tunnel}. In the tunneling regime, the semiclassical formula Eq.~\eqref{eq:tunnel} matches numerics when $a\approx0.25$. The semi-classical approximation reduces the computation time of the double quantum dot tunnel parameter by several orders of magnitude. For example computing the semi-classical Eq.~\eqref{eq:tunnel} typically takes about $11\unit{ms}$ on a basic core (Intel Core i5 CPU @ 1$\unit{GHz}$), while the calculations based on the two-dimensional Schr\"odinger equation discretized on a grid of size $80\times40$ takes about $5\unit{s}$.

\begin{figure}[h]
    \centering
    \includegraphics[width=0.49\textwidth]{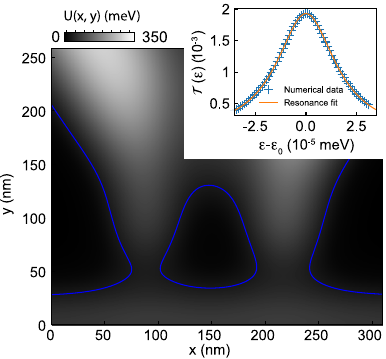}
    \caption{Applying the tunnel rate evaluations to a realistic device design. For the single quantum dot shown, with a resonance at energy $\varepsilon_0\approx 28.5\unit{meV}$ (shown as the blue contour), using the Landauer formula of Appendix~\ref{app:Landauer} we numerically find the tunnel rates between the dot and the reservoirs to be: $\Gamma_L\approx60\unit{MHz}$ and $\Gamma_R\approx 26\unit{kHz}$ (see inset). The semi-classical tunneling probabilities $\mathcal{T}_\text{barrier}$ for the left and right barriers evaluate to $5.2\times10^{-2}$ and $1.9\times10^{-5}$, respectively. This is consistent with a ratio $\Gamma/\mathcal{T}_{\rm barrier}=1.3\pm0.1\unit{GHz}$, a good agreement given the orders of magnitude between the tunneling rates of the left and right barriers. 
    The numerically computed single particle level spacing of the quantum dot is about $1.5\unit{meV}$, consistent with a dot radius $s_0\approx45\unit{nm}$. Here we assumed $m=0.05m_0$.}
    \label{fig:realistic_device}
\end{figure}

The system is set up as a central dot with two half dots, each a distance $d$ away from a central dot, and connected to infinite leads acting as reservoirs (Fig.~\ref{fig:realistic_device}). The tunnel rate to and from the reservoirs is obtained by fitting the resonance in the transmission probability, for the ground state of the quantum dot, following Eq\eqref{eq:low-temp-resonance}). The numerics in the inset of Fig.~\ref{fig:semiclassical_tunnel}(b) show a tunneling rate $\Gamma$ in good agreement with the semi-classical prediction. 

Regarding application to realistic devices, the tunneling parameters may be directly evaluated with the above semi-classical equations. If considering dot shapes, the parameters may be calculated by solving the two-dimensional Schrödinger equation for a given set of initial voltages as shown in Appendix~\ref{eq:tunnel}. The scaling with voltages can then be estimated using semi-classical equations. In Fig.~\ref{fig:realistic_device} we apply the method of Appendix~\ref{app:Landauer} to a realistic device design and compare with semi-classical methods. The correspondence is good, even with the significant asymmetry between the left and right barriers (Fig.~\ref{fig:realistic_device}). The extracted parameters are furthermore applicable to the calculation of transport properties in dot arrays based on master equations \cite{IhnBook, NazarovBook}. Such models can also serve as a basis for studying additional effects such as phonon-assisted transport \cite{Sowa2018,Gullans2018,craig2024differentiable} and other aspects of the semiconductor environment.

\section{Virtual Gate Methods Comparison}
\label{app:virtual_gate_comp}

\begin{figure*}[hbt!]
    \includegraphics[width=\textwidth]{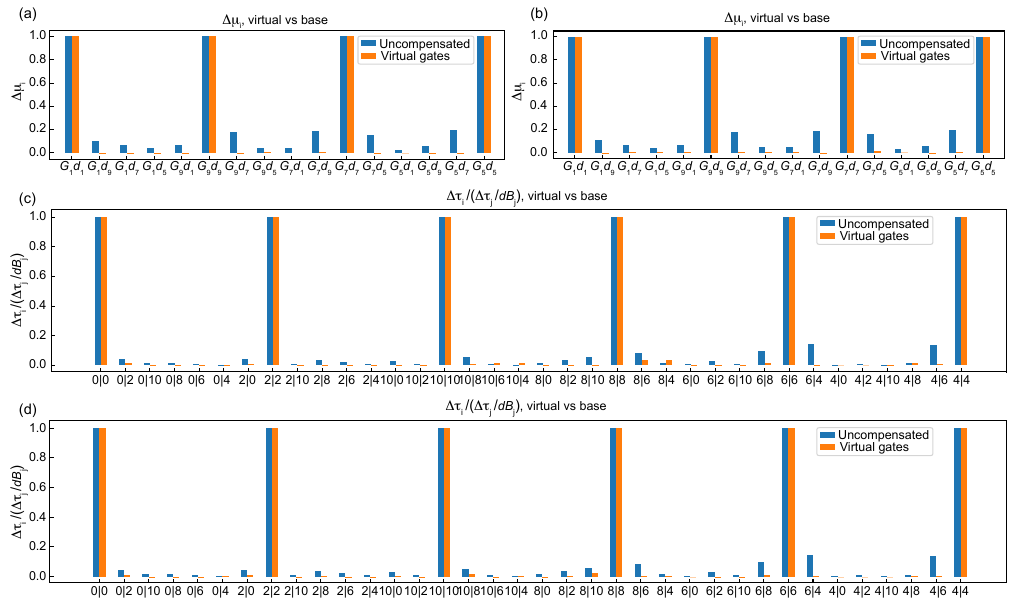}
    \caption{Observed difference in dot electrostatic potentials, $\mu_i$, and barrier tunneling couplings, $\tau_i$, normalized by their self-action terms ($\Delta \mu_i/P_i$ and $\Delta \tau_i/B_i$) when sweeping the physical gates (shown in blue) versus when sweeping virtual gates (shown in orange). In (a) and (b) the observed dot potential difference for dot $d_i$ (associated with the gate number of its plunger as in Fig.~\ref{fig:device_graph}(a)) is shown while sweeping gate (blue) or virtual gate (orange) $G_j$. Similarly, in (c) and (d) the barrier coupling difference of barrier $i$ is shown while sweeping gate $j$ displayed as $i|j$ (normalized by the self-action: 1/$\Delta \tau_j/dB_j$). This illustrates the subsystem crosstalk compensation using virtual gates derived from the submatrix approaches (a) and (c) on each of their respective compensation targets, i.e. the $\alpha$ submatrix for the dot potentials $\mu$ and the $\beta$ submatrix for the barrier couplings $\tau$. We observe that the compensation when sweeping the full crosstalk matrix inverted virtual gates as in (b) and (d) is better than the submatrix methods at compensating both the $\mu_i$ and $\tau_i$'s (as summarized in Table~\ref{table:crosstalk_mats_diff_from_I} of the main text), and additionally doesn't cause large changes in other parts of the system like the submatrix methods do when considering the full system (Fig.~\ref{fig:crosstalk_mats_comp}).}
    \label{fig:vg_comp_results}
\end{figure*}

\begin{figure*}[hbt!]
    \includegraphics[width=\textwidth]{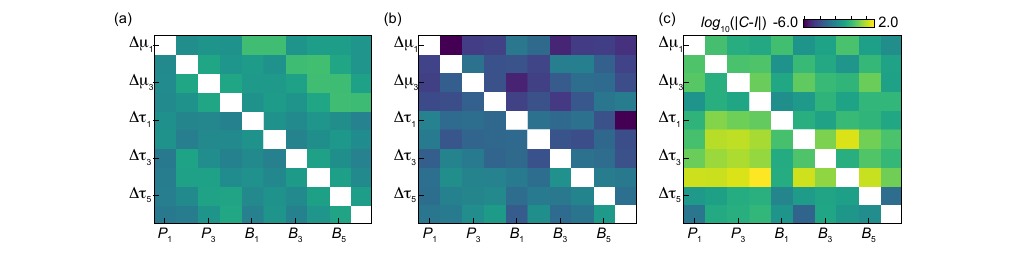}
    \caption{
    Logarithm of the absolute difference between the identity matrix and the observed crosstalk matrix when generated with voltage sweeps along (a) the physical gates, (b) virtual gates from the inverse of (a), and (c) virtual plungers and virtual barriers derived using pseudo-inverses of the $\alpha$ and $\beta$ submatrices, respectively. 
    }
    \label{fig:crosstalk_mats_comp}
\end{figure*}

In this section we compare virtual gate methodologies, which emphasize the importance of using the full crosstalk matrix to define virtual gates. We compare the full crosstalk matrix inverted virtual gates approach as outlined in the main text to previous methods used in the literature, in particular the methods of \cite{Volk2019} and \cite{Hsiao2020} where pseudo-inverses of the $\alpha$ and $\beta$ submatrices form the virtual plunger and barrier gates, respectively. We implement the aforementioned approaches using our graph-based simulator configured to model the same device as in the main text. We note that while the approach of \cite{Hsiao2020} includes virtual plungers of the $\alpha$ submatrix in the definition of the virtual barrier gates in order to minimize their effect on the dot electrochemical potential levels, in our simulation we found that sweeping this modified definition of the virtual barrier gates is accompanied by an increased change in the barrier tunnel couplings, which is undesirable for virtual barrier gates. We compare the efficacy of the crosstalk compensation methods by observing the average change in each dot electrochemical potential $\Delta \mu$ and each barrier tunnel coupling $\Delta \tau$ while varying the gate voltages along vectors defined by the physical gates, the full crosstalk matrix inverted virtual gates, and the combination of the pseudo-inverse virtual plungers and barriers. In Fig.~\ref{fig:vg_comp_results} we sweep along the pseudo-inverse virtual plungers (virtual barriers) while observing only the dots (barriers) and compare the total change in dot potential or barrier coupling to that when sweeping the full crosstalk matrix inverted virtual gates. These results are also summarized in Table~1 of the main text. In Fig.~\ref{fig:crosstalk_mats_comp} the comparison is over the full device and is an extension of Fig.~3 from the main text where the pseudo-inverse methods mentioned here are included.

\section{Graph Execution and Results} 
\label{app:graph_and_results}

In this section we give more details about how the digital surrogate graph computes target properties, its computational speed with different cpu and batch sizes, and the types of metrics used to choose deep learning models along with metrics of select models. To use the simulator, the user requests a list of names of the variables they want compute, conditioned on a dictionary of known values. For each name in the list, the graph checks if the variable is already known; if not, it will execute the node that has that name as output using inputs from the known value dictionary. If the inputs are not present in the dictionary, the graph recursively executes the required nodes, adding all intermediate variables to the known value dictionary. If the user has provided a set of known values that are sufficient to compute the desired targets, the graph will execute the minimal set of nodes required to return the desired result.
The system configuration holds data pertinent to all nodes, and a node configuration contains information relevant to a particular computation and its connectivity within the graph. 
To further improve efficiency, a recursive graph traversal algorithm ensures that only the necessary nodes are computed when inputting suitable pre-computed data.

\begin{table}[ht]
\centering
\begin{tabular}{|c|c|c|c|c|c|}
\hline
  & \multicolumn{5}{c|}{\textbf{Number of CPUs used}} \\
\cline{2-6}
\textbf{Batch Size}  & \textbf{20} & \textbf{10} & \textbf{5} & \textbf{2} & \textbf{1} \\
\hline
300 & 6.46 & 6.68 & 7.95 & 15.41 & 27.58 \\
150 & 6.46 & 6.62 & 8.26 & 15.31 & 27.74 \\
80 & 6.58 & 6.56 & 7.96 & 15.29 & 28.35 \\
40 & 6.7 & 6.73 & 8.01 & 15.25 & 27.74 \\
20 & 6.74 & 7.14 & 8.06 & 16.45 & 28.02 \\
10 & 7.6 & 7.35 & 8.13 & 17.7 & 28.24 \\
5 & 8.44 & 8.38 & 8.36 & 17.78 & 28.25 \\
2 & 16.39 & 16.39 & 16.07 & 15.72 & 28.19 \\
1 & 28.61 & 28.48 & 28.64 & 28.03 & 27.66 \\
\hline
\end{tabular}
\caption{Mean time taken (seconds) to compute dots with occupation using the physics-based graph.}
\label{table:slow_timings}
\end{table}

\begin{table}[ht]
\centering
\begin{tabular}{|c|c|c|c|c|c|}
\hline
  & \multicolumn{5}{c|}{\textbf{Number of CPUs used}} \\
\cline{2-6}

\textbf{Batch Size}  & \textbf{20} & \textbf{10} & \textbf{5} & \textbf{2} & \textbf{1} \\
\hline
300 & 0.12 & 0.11 & 0.11 & 0.15 & 0.22 \\
150 & 0.15 & 0.12 & 0.12 & 0.16 & 0.22 \\
80 & 0.2 & 0.15 & 0.13 & 0.17 & 0.22 \\
40 & 0.24 & 0.2 & 0.16 & 0.18 & 0.22 \\
20 & 0.34 & 0.26 & 0.23 & 0.21 & 0.22 \\
10 & 0.35 & 0.36 & 0.28 & 0.27 & 0.23 \\
5 & 0.42 & 0.4 & 0.45 & 0.42 & 0.25 \\
2 & 0.62 & 0.68 & 0.7 & 0.58 & 0.27 \\
1 & 0.44 & 0.5 & 0.44 & 0.43 & 0.29 \\
\hline
\end{tabular}
\caption{Mean time taken (seconds) to compute dots with occupation using the deep learning accelerated graph.}
\label{table:fast_timings}
\end{table}

\begin{table}[ht]
\centering
\begin{tabular}{|c|c|c|c|c|c|}
\hline

  & \multicolumn{5}{c|}{\textbf{Number of CPUs used}} \\
\cline{2-6}

\textbf{Batch Size}  & \textbf{20} & \textbf{10} & \textbf{5} & \textbf{2} & \textbf{1} \\
\hline
300 & 8.63 & 8.85 & 8.9 & 6.46 & 4.55 \\
150 & 6.88 & 8.17 & 8.28 & 6.37 & 4.61 \\
80 & 4.95 & 6.48 & 7.47 & 5.98 & 4.54 \\
40 & 4.18 & 5.1 & 6.17 & 5.6 & 4.48 \\
20 & 2.95 & 3.85 & 4.41 & 4.86 & 4.48 \\
10 & 2.83 & 2.82 & 3.56 & 3.76 & 4.3 \\
5 & 2.36 & 2.5 & 2.22 & 2.39 & 4.03 \\
2 & 1.62 & 1.48 & 1.42 & 1.73 & 3.71 \\
1 & 2.25 & 2.01 & 2.29 & 2.34 & 3.43 \\
\hline
\end{tabular}
\caption{Rate of computation (samples per second) of dots with occupation using the deep-learning accelerated graph.}
\label{table:fast_timings_rate}
\end{table}

\begin{table*}[hbt!]
\centering
\begin{tabular}{|c|c|c|c|c|c|c|c|c|}
\hline

  & \multicolumn{3}{c|}{\textbf{Near dot-forming voltages}} & \multicolumn{2}{c|}{\textbf{Uniform in V}}\\
\cline{2-6}
\textbf{run id} & Dot IoU  & 
SC pot MAE &  IoU classically-forbidden & SC pot MAE & IoU classically-forbidden\\
\hline
yfwFVY & 0.757 & 6.589 & \textbf{0.981} & \textbf{2.464} & \textbf{0.629} \\
oNGnVZ & 0.768 & 5.706 & 0.976 & 2.538 & 0.531 \\
RNvnAv & \textbf{0.795} & 5.851 & 0.978 & 2.796 & 0.506 \\
QUqLUJ & 0.789 & \textbf{5.51} & 0.976 & 2.562 & 0.565 \\
LuQIaz & 0.751 & 7.753 & 0.972 & 2.867 & 0.471 \\
OpyEVA & 0.778 & 6.14 & 0.975 & 2.683 & 0.513 \\
TmqTjD$^{\dag}$ & 0.769 & 6.267 & \textbf{0.981} & 2.605 & 0.583 \\
ySnPCa$^{\dag}$ & 0.716 & 6.558 & 0.976 & 2.713 & 0.556 \\

\hline
\end{tabular}
\caption{Evaluation metrics applied to trained deep learning models for gate voltages sampled from two distributions. Voltages are either sampled from a Gaussian distribution centered on a set of voltages that is known to produce quantum dots at all locations, or from a uniform distribution across the whole voltage space. Metrics are \textbf{1}. The intersection over union (IoU) of the area identified as dots \textbf{2}. The mean absolute error (MAE) in the self-consistent potential (mV) \textbf{3}. The intersection over union of regions determined to be classifically forbidden for charge carriers to access. The dot IoU is omitted from the Uniform in V results as dots are rarely formed when sampling from the entire voltage space uniformly. Bold results highlight the best model for each metric. Daggers indicate models which failed to find dot-forming voltages with automated ramp-up.}
\label{table:metrics}
\end{table*}

\end{document}